\documentclass[prl,reprint]{revtex4-1}
\pdfoutput=1
\usepackage{graphicx}
\usepackage{bm}
\usepackage{hyperref}
\usepackage{color}

\begin{document}

\title{Observation of reentrant quantum Hall states in the lowest Landau level}
\date{today}

\author{Yang Liu}
\affiliation{Department of Electrical Engineering,
Princeton University, Princeton, New Jersey 08544}
\author{C.G.\ Pappas}
\affiliation{Department of Electrical Engineering,
Princeton University, Princeton, New Jersey 08544}
\author{M.\ Shayegan}
\affiliation{Department of Electrical Engineering,
Princeton University, Princeton, New Jersey 08544}
\author{L.N.\ Pfeiffer}
\affiliation{Department of Electrical Engineering,
Princeton University, Princeton, New Jersey 08544}
\author{K.W.\ West}
\affiliation{Department of Electrical Engineering,
Princeton University, Princeton, New Jersey 08544}
\author{K.W.\ Baldwin}
\affiliation{Department of Electrical Engineering,
Princeton University, Princeton, New Jersey 08544}

\date{\today}

\begin{abstract}
  Measurements in very low disorder two-dimensional electrons confined
  to relatively wide GaAs quantum well samples with tunable density
  reveal reentrant $\nu=1$ integer quantum Hall states in the lowest
  Landau level near filling factors $\nu=4/5$ and 6/5. These states
  are not seen at low densities and become more prominent with
  increasing density and in wider wells. Our data suggest a close
  competition between different types of Wigner crystal states near
  these fillings. We also observe an intriguing disappearance and
  reemergence of the $\nu=4/5$ fractional quantum Hall effect with
  increasing density.
\end{abstract}


\maketitle

At low temperatures and subjected to a strong perpendicular magnetic
($B$), a low-disorder two-dimensional electron system (2DES) displays
a myriad of novel collective states \cite{Jain.CF.2007, Note1, Note2}
arising from the dominance of the Coulomb interaction energy over the
kinetic energy and the disorder potential. At high $B$, when the
electrons occupy the lowest ($N=0$) Landau level (LL) \cite{Note3},
the 2DES exhibits fractional quantum
Hall effect (FQHE) at several series of odd-denominator LL fractional
fillings $\nu=nh/eB$ ($n$ is the 2DES density) as it condenses into
incompressible liquid states \cite{Jain.CF.2007, Note1, Note2, Note3,
  Tsui.PRL.1982}. At even higher $B$, the last series of FQHE liquid
states is terminated by an insulating phase which is reentrant around
a FQHE at $\nu=1/5$ and extends to lower $\nu<1/5$
\cite{Jiang.PRL.1990, Goldman.PRL.1990, Note4}.
This insulating phase is generally believed to be an electron Wigner
crystal (WC), pinned by the small but ubiquitous disorder potential
\cite{Jiang.PRL.1990, Goldman.PRL.1990, Note4}.

In the higher ($N>0$) LLs, at the lowest temperatures and in the
cleanest samples, other collective states compete with the FQHE
states. These include anisotropic "stripe" phases at half-integer
fillings for $N\ge2$ and several insulating phases at non-integer
fillings which exhibit reentrant integer quantum Hall effect (RIQHE)
behavior \cite{Lilly.PRL.1999, Du.SSC.1999, Eisenstein.PRL.2002,
  Xia.PRL.2004}. The latter phases are most prominent in the second
($N=1$) LL, and their hallmark feature is a Hall resistance ($R_{xy}$)
which is quantized at the value of a nearby integer quantum Hall
plateau and is often accompanied by a vanishing longitudinal
resistance ($R_{xx}$) at the lowest achievable temperatures. The origin of these
RIQHE phases is not entirely clear, although they are widely
considered to signal the condensation of electrons in a
partially-filled LL into "bubble" phases where several electrons are
localized at one lattice site \cite{Lilly.PRL.1999, Du.SSC.1999,
  Eisenstein.PRL.2002, Xia.PRL.2004, Koulakov.PRL.1996,
  *Fogler.PRB.1997}. Such RIQHEs have not been seen until now in the
$N=0$ LL, consistent with the expected the instability of the bubble
phases in 2DESs in the lowest LL \cite{Note5}. 

We report here the observation of RIQHE in the lowest LL in very
clean, high-density 2DESs confined to relatively wide GaAs quantum
wells (QWs). Figure 1 highlights our main finding. It shows data taken
in a 42-nm-wide GaAs QW at two different densities. At the lower
density, the $R_{xx}$ and $R_{xy}$ traces show what is normally seen
in very clean 2DESs: strong QHE at $\nu=1$ and 2/3 and, between these
fillings, several FQHE states at $\nu=$ 4/5, 7/9, 8/11, and 5/7. At
the higher $n$, however, a RIQHE (marked by down arrows) is observed
near $1/\nu=1.20$, as evidenced by an $R_{xx}$ minimum and and
$R_{xy}$ quantized at $h/e^2$. Also evident is a developing RIQHE
state between $\nu=4/5$ and 7/9, signaled by a dip in $R_{xy}$ (up
arrow in Fig. 1(b)). As we detail below, the RIQHE phases near
$\nu=4/5$, as well as similar phases near $\nu=6/5$ on the low-field
flank of $\nu=1$, show a spectacular evolution with density. An
examination of the conditions under which we observe these reentrant
phases suggests that they are likely WC states, similar to those
observed near $\nu=1/5$ in high-quality 2DESs, and it is the larger
electron layer thickness in our samples that stabilizes them here in
the lowest LL near $\nu=1$.

\begin{figure}[!b]
\includegraphics[width=0.45\textwidth]{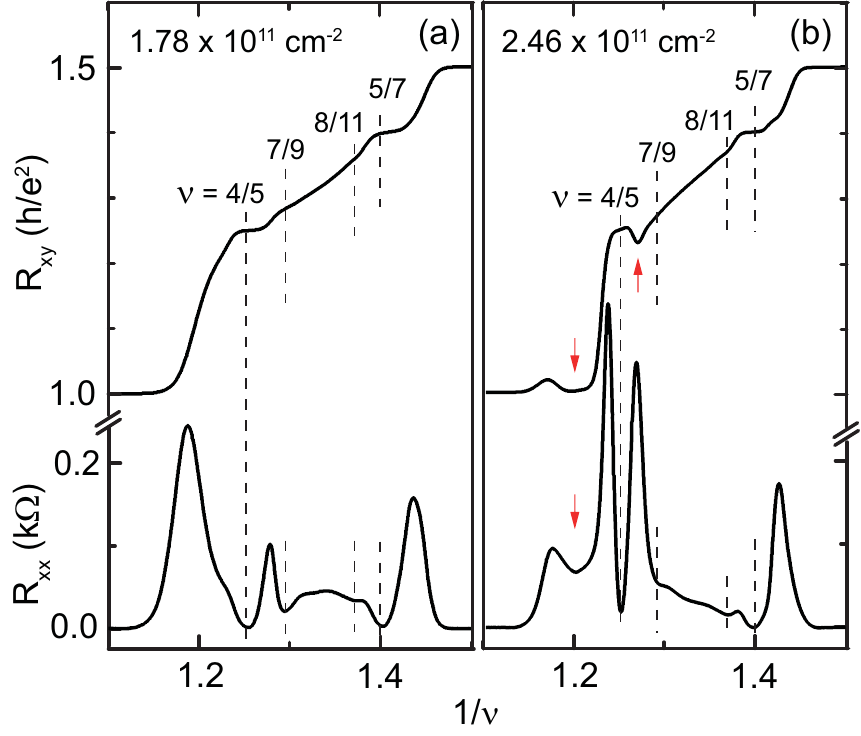}
\caption{\label{fig:highlight} ${R_{xx}}$ and $R_{xy}$ vs. $1/\nu$
  traces at $T=30$ mK for a 42-nm-wide GaAs QW at two densities: (a)
  $n=1.78$, and (b) $2.46\times 10^{11}$ cm${^{-2}}$. In (b) the RIQHE
  phases observed on two sides of $\nu=4/5$ are marked by arrows. Note
  also the two sharp $R_{xx}$ maxima surrounding the $\nu=4/5$
  $R_{xx}$ minimum.}
\end{figure}

\begin{figure}[tbp]
\includegraphics[width=0.45\textwidth]{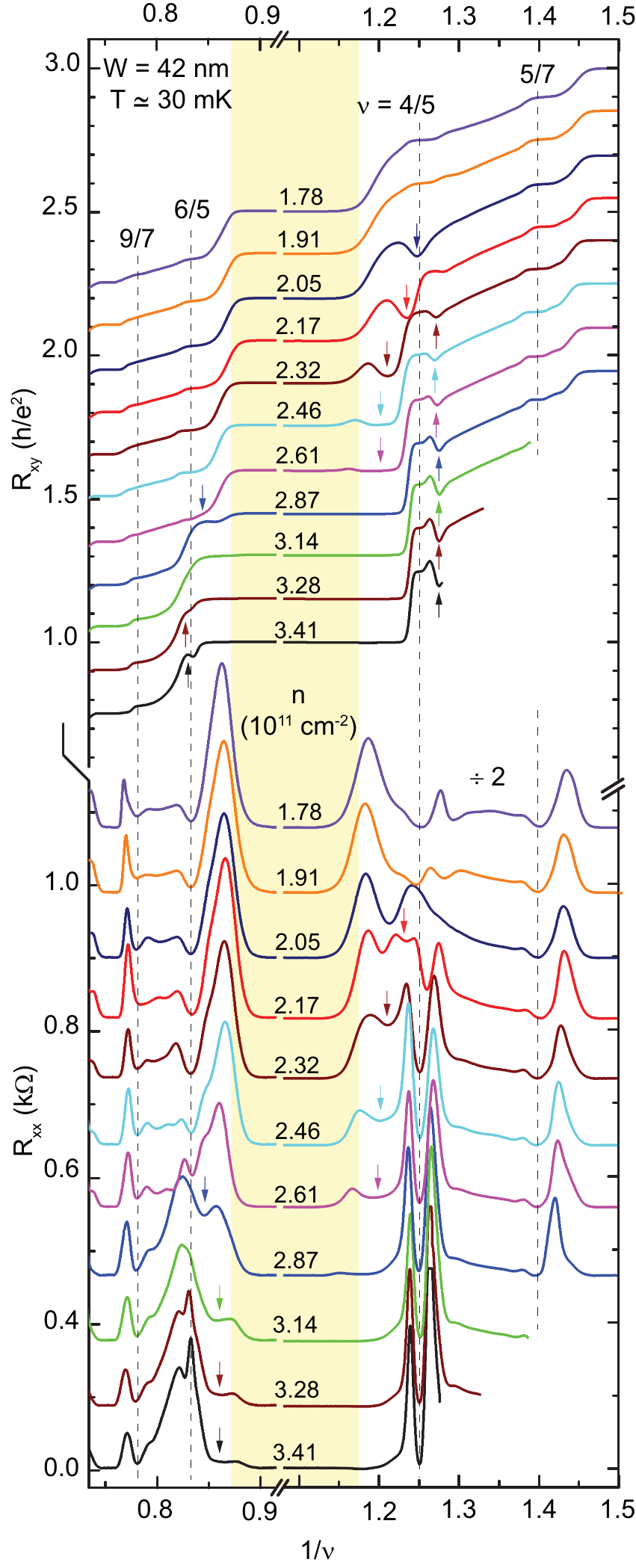}
\caption{\label{fig:42nm}(color online) Waterfall plots of ${R_{xx}}$
  and ${R_{xy}}$ vs. $1/\nu$ for the 42-nm-wide QW as $n$ is changed
  from $1.78$ to $3.41\times 10^{11}$ cm${^{-2}}$. Traces are shifted
  vertically for clarity. The down arrows mark the development of
  RIQHE phases near $1/\nu=1.20$ as $n$ is raised from $2.05$ to
  $2.61\times 10^{11}$ cm${^{-2}}$, and near $1/\nu=0.86$ as $n$ is
  further increased. The up arrows in the top panel mark the
  development of similar, albeit weaker, RIQHE phases near
  $1/\nu=1.28$ and 0.83. The yellow band marks the range
  $0.85\le\nu\le 1.15$; see text.}
\end{figure}

Our samples were grown by molecular beam epitaxy, and each consist of
a wide GaAs quantum well (QW) bounded on each side by undoped
Al$_{0.24}$Ga$_{0.76}$As spacer layers and Si $\delta$-doped
layers. We report here data for three samples, with QW widths $W=$ 31,
42 and 44 nm, and as-grown densities of $n\simeq$ 3.3, 2.5 and 3.8
$\times 10^{11}$ cm$^{-2}$, respectively. The low-temperature
mobilities of these samples are $\mu \simeq$ 670, 910 and 600
m$^2$/Vs, respectively. The samples have a van der Pauw geometry and
each is fitted with an evaporated Ti/Au front-gate and an In
back-gate. We carefully control $n$ and the charge distribution
symmetry in the QW by applying voltage biases to these gates
\cite{Suen.PRL.1994, Liu.PRB.2011}. All the data reported here were
taken by adjusting the front- and back-gate biases so that the total
charge distribution is symmetric. The measurements were carried out in
superconducting and resistive magnets with maximum fields of 18 T and
35 T respectively. We used low-frequency ($\simeq$ 10 Hz) lock-in
techniques to measure the transport coefficients.

Figure \ref{fig:42nm} shows a series of $R_{xx}$ and $R_{xy}$ traces
in the range $2/3 < \nu < 4/3$ for the 42-nm-wide QW sample, taken as
$n$ is changed from 1.78 to $3.41\times 10^{11}$cm$^{-2}$. These
traces reveal a remarkable evolution for the different reentrant
phases of this 2DES. At the lowest $n$, FQHE states are seen at $\nu=$
9/7, 6/5, 4/5, and 5/7. When $n$ is increased to $2.05\times 10^{11}$
cm$^{-2}$, the $\nu=4/5$ FQHE disappears and a minimum in $R_{xy}$
develops near $\nu=4/5$. As $n$ is further increased, the $R_{xy}$
minimum becomes deeper and moves towards $1/\nu=1.20$, and an $R_{xx}$
minimum starts to develop at the same filling (see down arrows near
$1/\nu=1.20$ in Fig. 2). Meanwhile, the $\nu=4/5$ FQHE reappears to
the right of these minima. As we keep increasing $n$, the $R_{xy}$
minimum deepens and becomes quantized at $h/e^2$ for $n\ge2.46\times
10^{11}$ cm$^{-2}$, and the $R_{xx}$ minimum gets deeper and vanishes
for $n>2.87\times 10^{11}$ cm$^{-2}$. At the highest $n$ these minima
merge into the $R_{xy}$ plateau and the $R_{xx}$ minimum near $\nu=1$.

Figure 2 traces also show that, with increasing $n$, another $R_{xy}$
minimum starts to develop on the right side of $\nu=4/5$, as marked by
the up arrows at $1/\nu=1.28$. This minimum, too, becomes deeper at
higher $n$ and, as we will show later, approaches $h/e^2$. Also
noteworthy are the two sharp $R_{xx}$ maxima on the flanks
of $\nu=4/5$ after the $\nu=4/5$ FQHE reemerges at high densities.

The evolution observed on the right side of $\nu=1$ is qualitatively
seen on the left side also, but at higher $n$. The $\nu=6/5$
FQHE, e.g., disappears at $n=2.87\times 10^{11}$ cm$^{-2}$ and a dip
in $R_{xx}$ develops near $1/\nu=0.86$ (see down arrows in Fig. 2),
concomitant with $R_{xy}$ lifting up and eventually becoming quantized
at $h/e^2$.


\begin{figure}
\includegraphics[width=0.45\textwidth]{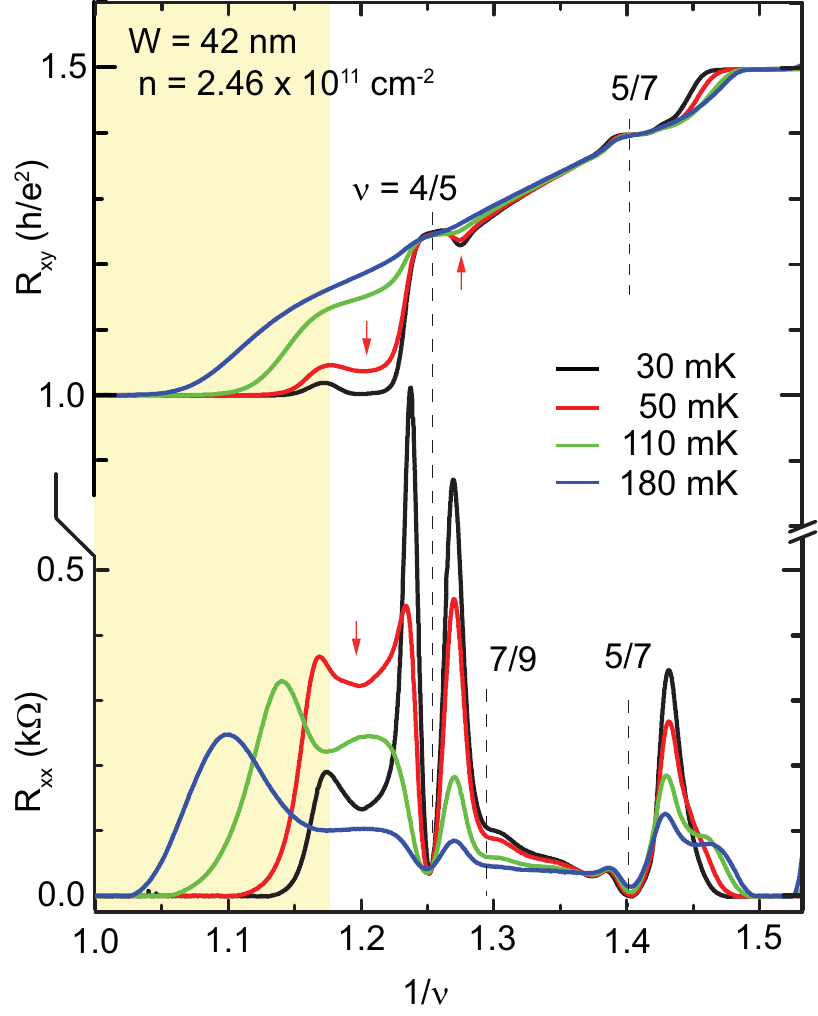}
\caption{\label{fig:tempdep}(color online) Evolution of ${R_{xx}}$ and
  ${R_{xy}}$ with temperature for the 42-nm QW at
  $n=2.46\times 10^{11}$ cm${^{-2}}$.}
\end{figure}


Figure~\ref{fig:tempdep} illustrates the $T$-dependence of $R_{xx}$
and $R_{xy}$ at $n=2.46\times10^{11}$ cm$^{-2}$. At the highest $T$,
the $R_{xx}$ and $R_{xy}$ traces near $\nu=4/5$ look "normal": There
is a relatively strong FQHE at $\nu=4/5$ as signaled by a deep minimum
in $R_{xx}$ and an $R_{xy}$ plateau at $5h/4e^2$. Away from $\nu=4/5$,
$R_{xy}$ follows a nearly linear dependence on $B$. As $T$ is lowered,
however, $R_{xy}$ develops a minimum near $1/\nu=1.20$ which
eventually turns to a plateau quantized at $h/e^2$ at the lowest
$T$. Meanwhile, a strong minimum develops in $R_{xx}$ at
$1/\nu=1.20$. On the right side of $\nu=4/5$, another $R_{xy}$ minimum
is developing.


Data taken on the 31- and 44-nm-wide QW samples reveal the generality
of these phenomena. Figure 4(a) shows data for the 31-nm QW. Similar
to the data of Fig. 2, as $n$ is increased, the $\nu=4/5$ FQHE is
quickly destroyed and $R_{xy}$ develops a deep minimum which
approaches the $\nu=1$ plateau at $h/e^2$ near the highest $B$ that we
can achieve in this sample. The resemblance of the traces shown in
Fig. 4(a) to the top four $R_{xy}$ and $R_{xx}$ traces in Fig. 2 is
clear. Note, however, that in the narrower QW of Fig. 4(a) we need
much higher $n$ to reproduce what is seen in the wider QW of Fig. 2.

\begin{figure}
\includegraphics[width=0.48\textwidth]{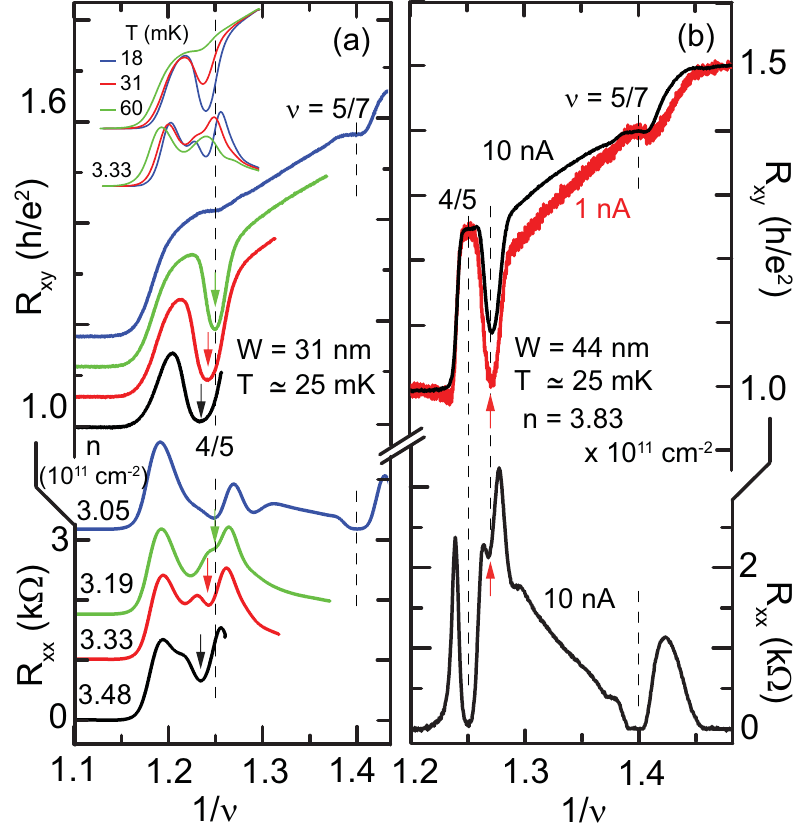}
\caption{\label{fig:31nm}(color online) (a) Waterfall plots of
  $R_{xx}$ and $R_{xy}$ vs. $1/\nu$ for the 31-nm QW sample,
  showing how the RIQHE phase near $1/\nu=1.20$ starts to develop as
  $n$ is raised from $3.05$ to $3.48\times 10^{11}$ cm${^{-2}}$. Data
  are shifted vertically for clarity. The inset shows the
  $T$-dependence of the RIQHE at $n=3.33\times 10^{11}$ cm${^{-2}}$. (b) $R_{xx}$ and $R_{xy}$ traces for
  the 44-nm QW at $n=3.83\times 10^{11}$ cm${^{-2}}$ displaying a
  RIQHE phase at $1/\nu=1.28$.}
\end{figure}

In Fig. 4(b) we show data for the 44-nm QW at a very high density
($n=3.83\times 10^{11}$ cm${^{-2}}$). Given the larger QW width and
higher $n$ of this sample compared to Fig. 2 sample, we expect the
RIQHE near $1/\nu=1.20$ to be fully developed, and the RIQHE on the
high-field side of $\nu=4/5$ whose emergence is hinted at in Fig. 2
$R_{xy}$ traces (see up arrows near $1/\nu=1.28$ in Fig. 2) to become
more pronounced. This is indeed seen in Fig. 4(b): The $R_{xy}$ trace
shows a very deep minimum on the high-field side of $\nu=4/5$ at
$1/\nu=1.28$, nearly reaching the $\nu=1$ plateau at $h/e^2$ when the
smallest sample current (1 nA) is used to achieve the lowest electron
temperature for this sample. Note also the appearance of a small but
clearly visible $R_{xx}$ minimum at $1/\nu=1.28$, consistent with the
development of RIQHE at this filling \cite{Note6}.

A natural interpretation of the RIQHE phases that we observe near
$\nu=4/5$ and $6/5$ is that these are pinned WC phases, similar to the
insulating phases seen around the $\nu=1/5$ FQHE and extending to
$\nu\ll 1/5$ \cite{Jiang.PRL.1990,Goldman.PRL.1990,Note4}. In the
present case, electrons at $\nu=1\pm\nu^*$ can be considered as a
filled LL, which is inert, plus excess electrons/holes with filling
factor $\nu^*$, which could conduct. At sufficiently small values of
$\nu^*$ and low temperatures, these excess electrons/holes crystalize
into a solid phase which is pinned by disorder and does not
participate in transport. Thus the magneto-transport coefficients,
$R_{xx}$ and $R_{xy}$, approach those of the $\nu=1$ IQHE.

In high-quality 2DESs, there have been no prior reports of RIQHE phases
near $\nu=1$ in the lowest LL from low-frequency (essentially dc)
magneto-transport measurements\cite{Note7}.
Our results indicate that such RIQHE phases set in only at high
densities and in relatively wide QWs. We suggest that it is the large
electron layer thickness (spread of the electron wavefunction
perpendicular to the 2D plane) in our wide QW samples that induces the
WC formation \cite{Note8}.
Supporting this conjecture, theoretical work \cite{Price.PRB.1995}
indeed indicates that in wide QWs, thanks to the softening of the
Coulomb interaction at short distances, a WC phase is favored over the
FQHE liquid state if the QW width becomes several times larger than
the magnetic length ($l_B$). In our samples, $W/l_B$ is large and
ranges from $\simeq$4.6 to $\simeq$ 6.0 at the densities above which
we start to observe the RIQHE near $\nu=4/5$, qualitatively consistent
with the theoretical results.

In a relevant study, high-frequency (microwave) resonances were
observed very close to $\nu=1$ ($\nu^*<0.15$) in high-quality 2DESs
and were interpreted as signatures of a pinned WC
\cite{Chen.PRL.2003}. We have highlighted this filling range with a
yellow band in Figs. 2 and 3. This is the range where we see a
deep $R_{xx}$ minimum and a very well quantized $R_{xy}$ (at $h/e^2$),
signaling a strong $\nu=1$ QHE. In our samples, however, we observe
RIQHE phases at relatively large values of $\nu^*$, extending to
$\nu^*>1/5$. In particular, $1/\nu\simeq1.28$ where we see the RIQHE
on the right side of $\nu=4/5$ corresponds to $\nu^*\simeq0.22$. This
is clearly outside the $\nu^*$ range where Chen $et$ $al.$ observed
microwave resonances, indicating that the RIQHE phases we are
reporting here are distinct from the phase documented in
Ref. \cite{Chen.PRL.2003}. Also the RIQHE we observe at $1/\nu=1.20$
($\nu^*\simeq0.17$) is separated from the deep QHE region very near
$\nu=1$ by maxima in $R_{xx}$ and $R_{xy}$ (see, e.g., the
right-hand-side edge of the yellow band in Figs. 2 and 3). This observation
provides additional evidence that two distinct insulating phases
exist. If so, then these resistance maxima signal additional scattering
at the domain walls separating these phases.
 
It is worth noting that microwave experiments at very small $\nu$ have
revealed two distinct resonances \cite{Chen.PRL.2004}. One resonance
("$A$") was seen in the insulating phases reentrant around $\nu=1/5$,
and another ("$B$") was dominant at very small fillings
($\nu<0.15$). It was suggested that these resonances signal the
existence of two different types of correlated WCs, stabilized by the
crystallization of composite Fermions with different number of flux
quanta attached to them. Such an interpretation is corroborated by
theoretical calculations \cite{Yi.PRB.1998, *Narevich.PRB.2001,
  Chang.PRL.2005, *Chang.PRB.2006}. It is tempting to associate the
RIQHE phases we observe reentrant around $\nu=4/5$ ($\nu^*=1/5$) with
the type "\emph{A}" WC and the resonance seen very near $\nu=1$
($\nu^*<0.15$) in Ref. \cite{Chen.PRL.2003} with the type "\emph{B}"
WC.

A very intriguing aspect of our data is the disappearance followed by
a rapid reemergence of the $\nu=4/5$ ($\nu^*=1/5$) FQHE when the RIQHE
near $\nu=4/5$ starts to be seen \cite{Note9}.
We do not have a clear explanation for this observation. We speculate
that it might signal the existence of multiple types of WC phases that
have ground-state energies very close to the ground-state energy of
the FQHE liquid phase. Theoretical calculations indeed indicate that
such WC phases, which are based on composite Fermion FQHE liquid
wavefunctions, do exist and might well describe the insulating phases
observed near $\nu=1/5$ \cite{Yi.PRB.1998, *Narevich.PRB.2001,
  Chang.PRL.2005, *Chang.PRB.2006}. The disappearance and
reappearance of the $\nu=4/5$ FQHE we observe may stem from a close
competition between the FQHE and such WC states. Consistent with this
scenario is the following observation in Fig. 4(a). When the $\nu=4/5$
FQHE disappears, $R_{xy}$ at $\nu=4/5$ immediately starts to dip down
towards $h/e^2$, suggesting a transition to a RIQHE phase. This is in
sharp contrast to $R_{xy}$ maintaining its value of $5h/4e^2$ on the
classical Hall line, if the $\nu=4/5$ FQHE state made a transition to
a compressible liquid phase.

In conclusion, we observe RIQHE phases near $\nu=4/5$ and 6/5 in the
lowest LL in very clean 2DESs confined to relatively wide GaAs QWs. In
a given QW, the RIQHE is absent at low densities and develops above a
certain density which is higher for narrower QWs. Our observations are
consistent with the crystallization of excess electrons/holes in the
unfilled lowest LL into a WC.

\begin{acknowledgments}
  We thank L. W. Engel and J. K. Jain for illuminating discussions. We
  acknowledge support through the Moore Foundation and the NSF
  (DMR-0904117 and MRSEC DMR-0819860) for sample fabrication and
  characterization, and the DOE BES (DE-FG0200-ER45841) for
  measurements. This work was performed at the National High Magnetic
  Field Laboratory, which is supported by NSF Cooperative Agreement
  No. DMR-0654118, by the State of Florida, and by the DOE. We thank
  L. W. Engel for illuminating discussions, and E. Palm, J. H. Park,
  T. P. Murphy and G. E. Jones for technical assistance.
\end{acknowledgments}

\bibliography{../bib_full}

\begin{thebibliography}{28}%
\makeatletter
\providecommand \@ifxundefined [1]{%
 \@ifx{#1\undefined}
}%
\providecommand \@ifnum [1]{%
 \ifnum #1\expandafter \@firstoftwo
 \else \expandafter \@secondoftwo
 \fi
}%
\providecommand \@ifx [1]{%
 \ifx #1\expandafter \@firstoftwo
 \else \expandafter \@secondoftwo
 \fi
}%
\providecommand \natexlab [1]{#1}%
\providecommand \enquote  [1]{``#1''}%
\providecommand \bibnamefont  [1]{#1}%
\providecommand \bibfnamefont [1]{#1}%
\providecommand \citenamefont [1]{#1}%
\providecommand \href@noop [0]{\@secondoftwo}%
\providecommand \href [0]{\begingroup \@sanitize@url \@href}%
\providecommand \@href[1]{\@@startlink{#1}\@@href}%
\providecommand \@@href[1]{\endgroup#1\@@endlink}%
\providecommand \@sanitize@url [0]{\catcode `\\12\catcode `\$12\catcode
  `\&12\catcode `\#12\catcode `\^12\catcode `\_12\catcode `\%12\relax}%
\providecommand \@@startlink[1]{}%
\providecommand \@@endlink[0]{}%
\providecommand \url  [0]{\begingroup\@sanitize@url \@url }%
\providecommand \@url [1]{\endgroup\@href {#1}{\urlprefix }}%
\providecommand \urlprefix  [0]{URL }%
\providecommand \Eprint [0]{\href }%
\@ifxundefined \urlstyle {%
  \providecommand \doi  [0]{\begingroup \@sanitize@url \@doi}%
  \providecommand \@doi [1]{\endgroup \@@startlink {\doibase
  #1}doi:\discretionary {}{}{}#1\@@endlink }%
}{%
  \providecommand \doi  [0]{doi:\discretionary{}{}{}\begingroup
  \urlstyle{rm}\Url }%
}%
\providecommand \doibase [0]{http://dx.doi.org/}%
\providecommand \Doi [0]{\begingroup \@sanitize@url \@Doi }%
\providecommand \@Doi  [1]{\endgroup\@@startlink{\doibase#1}\@@Doi}%
\providecommand \@@Doi [1]{#1\@@endlink}%
\providecommand \selectlanguage [0]{\@gobble}%
\providecommand \bibinfo  [0]{\@secondoftwo}%
\providecommand \bibfield  [0]{\@secondoftwo}%
\providecommand \translation [1]{[#1]}%
\providecommand \BibitemOpen [0]{}%
\providecommand \bibitemStop [0]{}%
\providecommand \bibitemNoStop [0]{.\EOS\space}%
\providecommand \EOS [0]{\spacefactor3000\relax}%
\providecommand \BibitemShut  [1]{\csname bibitem#1\endcsname}%
\bibitem [{\citenamefont {Jain}(2007)}]{Jain.CF.2007}%
  \BibitemOpen
  \bibfield  {author} {\bibinfo {author} {\bibfnamefont {J.~K.}\ \bibnamefont
  {Jain}},\ }\href@noop {} {\emph {\bibinfo {title} {{Composite Fermions}}}}\
  (\bibinfo  {publisher} {Cambridge University Press, Cambridge, UK},\ \bibinfo
  {year} {2007})\BibitemShut {NoStop}%
\bibitem [{Note1()}]{Note1}%
  \BibitemOpen
  \bibinfo {note} {\protect \emph {Perspectives in Quantum Hall Effects},
  Edited by S. Das Sarma and A. Pinczuk (Wiley, New York, 1997).}\BibitemShut
  {Stop}%
\bibitem [{Note2()}]{Note2}%
  \BibitemOpen
  \bibinfo {note} {See, e.g., M. Shayegan, ''Flatland Electrons in High
  Magnetic Fields,'' in \protect \emph {High Magnetic Fields: Science and
  Technology}, Vol. 3, edited by Fritz Herlach and Noboru Miura (World
  Scientific, Singapore, 2006), pp. 31-60. [cond-mat/0505520]}\BibitemShut
  {NoStop}%
\bibitem [{Note3()}]{Note3}%
  \BibitemOpen
  \bibinfo {note} {This regime corresponds to $\nu <2$, accounting for the two
  $N=0$ LLs, one for each spin orientation.}\BibitemShut {Stop}%
\bibitem [{\citenamefont {Tsui}\ \emph {et~al.}(1982)\citenamefont {Tsui},
  \citenamefont {Stormer},\ and\ \citenamefont {Gossard}}]{Tsui.PRL.1982}%
  \BibitemOpen
  \bibfield  {author} {\bibinfo {author} {\bibfnamefont {D.~C.}\ \bibnamefont
  {Tsui}}, \bibinfo {author} {\bibfnamefont {H.~L.}\ \bibnamefont {Stormer}}, \
  and\ \bibinfo {author} {\bibfnamefont {A.~C.}\ \bibnamefont {Gossard}},\
  }\Doi {10.1103/PhysRevLett.48.1559} {\bibfield  {journal} {\bibinfo
  {journal} {Phys. Rev. Lett.},\ }\textbf {\bibinfo {volume} {48}},\ \bibinfo
  {pages} {1559} (\bibinfo {year} {1982})}\BibitemShut {NoStop}%
\bibitem [{\citenamefont {Jiang}\ \emph {et~al.}(1990)\citenamefont {Jiang},
  \citenamefont {Willett}, \citenamefont {Stormer}, \citenamefont {Tsui},
  \citenamefont {Pfeiffer},\ and\ \citenamefont {West}}]{Jiang.PRL.1990}%
  \BibitemOpen
  \bibfield  {author} {\bibinfo {author} {\bibfnamefont {H.~W.}\ \bibnamefont
  {Jiang}}, \bibinfo {author} {\bibfnamefont {R.~L.}\ \bibnamefont {Willett}},
  \bibinfo {author} {\bibfnamefont {H.~L.}\ \bibnamefont {Stormer}}, \bibinfo
  {author} {\bibfnamefont {D.~C.}\ \bibnamefont {Tsui}}, \bibinfo {author}
  {\bibfnamefont {L.~N.}\ \bibnamefont {Pfeiffer}}, \ and\ \bibinfo {author}
  {\bibfnamefont {K.~W.}\ \bibnamefont {West}},\ }\Doi
  {10.1103/PhysRevLett.65.633} {\bibfield  {journal} {\bibinfo  {journal}
  {Phys. Rev. Lett.},\ }\textbf {\bibinfo {volume} {65}},\ \bibinfo {pages}
  {633} (\bibinfo {year} {1990})}\BibitemShut {NoStop}%
\bibitem [{\citenamefont {Goldman}\ \emph {et~al.}(1990)\citenamefont
  {Goldman}, \citenamefont {Santos}, \citenamefont {Shayegan},\ and\
  \citenamefont {Cunningham}}]{Goldman.PRL.1990}%
  \BibitemOpen
  \bibfield  {author} {\bibinfo {author} {\bibfnamefont {V.~J.}\ \bibnamefont
  {Goldman}}, \bibinfo {author} {\bibfnamefont {M.}~\bibnamefont {Santos}},
  \bibinfo {author} {\bibfnamefont {M.}~\bibnamefont {Shayegan}}, \ and\
  \bibinfo {author} {\bibfnamefont {J.~E.}\ \bibnamefont {Cunningham}},\ }\Doi
  {10.1103/PhysRevLett.65.2189} {\bibfield  {journal} {\bibinfo  {journal}
  {Phys. Rev. Lett.},\ }\textbf {\bibinfo {volume} {65}},\ \bibinfo {pages}
  {2189} (\bibinfo {year} {1990})}\BibitemShut {NoStop}%
\bibitem [{Note4()}]{Note4}%
  \BibitemOpen
  \bibinfo {note} {See articles by H. A. Fertig and by M. Shayegan in Ref.
  \cite {Note1}}\BibitemShut {NoStop}%
\bibitem [{\citenamefont {Lilly}\ \emph {et~al.}(1999)\citenamefont {Lilly},
  \citenamefont {Cooper}, \citenamefont {Eisenstein}, \citenamefont
  {Pfeiffer},\ and\ \citenamefont {West}}]{Lilly.PRL.1999}%
  \BibitemOpen
  \bibfield  {author} {\bibinfo {author} {\bibfnamefont {M.~P.}\ \bibnamefont
  {Lilly}}, \bibinfo {author} {\bibfnamefont {K.~B.}\ \bibnamefont {Cooper}},
  \bibinfo {author} {\bibfnamefont {J.~P.}\ \bibnamefont {Eisenstein}},
  \bibinfo {author} {\bibfnamefont {L.~N.}\ \bibnamefont {Pfeiffer}}, \ and\
  \bibinfo {author} {\bibfnamefont {K.~W.}\ \bibnamefont {West}},\ }\Doi
  {10.1103/PhysRevLett.82.394} {\bibfield  {journal} {\bibinfo  {journal}
  {Phys. Rev. Lett.},\ }\textbf {\bibinfo {volume} {82}},\ \bibinfo {pages}
  {394} (\bibinfo {year} {1999})}\BibitemShut {NoStop}%
\bibitem [{\citenamefont {Du}\ \emph {et~al.}(1999)\citenamefont {Du},
  \citenamefont {Tsui}, \citenamefont {Stormer}, \citenamefont {Pfeiffer},
  \citenamefont {Baldwin},\ and\ \citenamefont {West}}]{Du.SSC.1999}%
  \BibitemOpen
  \bibfield  {author} {\bibinfo {author} {\bibfnamefont {R.~R.}\ \bibnamefont
  {Du}}, \bibinfo {author} {\bibfnamefont {D.~C.}\ \bibnamefont {Tsui}},
  \bibinfo {author} {\bibfnamefont {H.~L.}\ \bibnamefont {Stormer}}, \bibinfo
  {author} {\bibfnamefont {L.~N.}\ \bibnamefont {Pfeiffer}}, \bibinfo {author}
  {\bibfnamefont {K.~W.}\ \bibnamefont {Baldwin}}, \ and\ \bibinfo {author}
  {\bibfnamefont {K.~W.}\ \bibnamefont {West}},\ }\Doi {DOI:
  10.1016/S0038-1098(98)00578-X} {\bibfield  {journal} {\bibinfo  {journal}
  {Solid State Communications},\ }\textbf {\bibinfo {volume} {109}},\ \bibinfo
  {pages} {389 } (\bibinfo {year} {1999})}\BibitemShut {NoStop}%
\bibitem [{\citenamefont {Eisenstein}\ \emph {et~al.}(2002)\citenamefont
  {Eisenstein}, \citenamefont {Cooper}, \citenamefont {Pfeiffer},\ and\
  \citenamefont {West}}]{Eisenstein.PRL.2002}%
  \BibitemOpen
  \bibfield  {author} {\bibinfo {author} {\bibfnamefont {J.~P.}\ \bibnamefont
  {Eisenstein}}, \bibinfo {author} {\bibfnamefont {K.~B.}\ \bibnamefont
  {Cooper}}, \bibinfo {author} {\bibfnamefont {L.~N.}\ \bibnamefont
  {Pfeiffer}}, \ and\ \bibinfo {author} {\bibfnamefont {K.~W.}\ \bibnamefont
  {West}},\ }\Doi {10.1103/PhysRevLett.88.076801} {\bibfield  {journal}
  {\bibinfo  {journal} {Phys. Rev. Lett.},\ }\textbf {\bibinfo {volume} {88}},\
  \bibinfo {pages} {076801} (\bibinfo {year} {2002})}\BibitemShut {NoStop}%
\bibitem [{\citenamefont {Xia}\ \emph {et~al.}(2004)\citenamefont {Xia},
  \citenamefont {Pan}, \citenamefont {Vicente}, \citenamefont {Adams},
  \citenamefont {Sullivan}, \citenamefont {Stormer}, \citenamefont {Tsui},
  \citenamefont {Pfeiffer}, \citenamefont {Baldwin},\ and\ \citenamefont
  {West}}]{Xia.PRL.2004}%
  \BibitemOpen
  \bibfield  {author} {\bibinfo {author} {\bibfnamefont {J.~S.}\ \bibnamefont
  {Xia}}, \bibinfo {author} {\bibfnamefont {W.}~\bibnamefont {Pan}}, \bibinfo
  {author} {\bibfnamefont {C.~L.}\ \bibnamefont {Vicente}}, \bibinfo {author}
  {\bibfnamefont {E.~D.}\ \bibnamefont {Adams}}, \bibinfo {author}
  {\bibfnamefont {N.~S.}\ \bibnamefont {Sullivan}}, \bibinfo {author}
  {\bibfnamefont {H.~L.}\ \bibnamefont {Stormer}}, \bibinfo {author}
  {\bibfnamefont {D.~C.}\ \bibnamefont {Tsui}}, \bibinfo {author}
  {\bibfnamefont {L.~N.}\ \bibnamefont {Pfeiffer}}, \bibinfo {author}
  {\bibfnamefont {K.~W.}\ \bibnamefont {Baldwin}}, \ and\ \bibinfo {author}
  {\bibfnamefont {K.~W.}\ \bibnamefont {West}},\ }\Doi
  {10.1103/PhysRevLett.93.176809} {\bibfield  {journal} {\bibinfo  {journal}
  {Phys. Rev. Lett.},\ }\textbf {\bibinfo {volume} {93}},\ \bibinfo {pages}
  {176809} (\bibinfo {year} {2004})}\BibitemShut {NoStop}%
\bibitem [{\citenamefont {Koulakov}\ \emph {et~al.}(1996)\citenamefont
  {Koulakov}, \citenamefont {Fogler},\ and\ \citenamefont
  {Shklovskii}}]{Koulakov.PRL.1996}%
  \BibitemOpen
  \bibfield  {author} {\bibinfo {author} {\bibfnamefont {A.~A.}\ \bibnamefont
  {Koulakov}}, \bibinfo {author} {\bibfnamefont {M.~M.}\ \bibnamefont
  {Fogler}}, \ and\ \bibinfo {author} {\bibfnamefont {B.~I.}\ \bibnamefont
  {Shklovskii}},\ }\Doi {10.1103/PhysRevLett.76.499} {\bibfield  {journal}
  {\bibinfo  {journal} {Phys. Rev. Lett.},\ }\textbf {\bibinfo {volume} {76}},\
  \bibinfo {pages} {499} (\bibinfo {year} {1996})}\BibitemShut {NoStop}%
\bibitem [{\citenamefont {Fogler}\ and\ \citenamefont
  {Koulakov}(1997)}]{Fogler.PRB.1997}%
  \BibitemOpen
  \bibfield  {author} {\bibinfo {author} {\bibfnamefont {M.~M.}\ \bibnamefont
  {Fogler}}\ and\ \bibinfo {author} {\bibfnamefont {A.~A.}\ \bibnamefont
  {Koulakov}},\ }\Doi {10.1103/PhysRevB.55.9326} {\bibfield  {journal}
  {\bibinfo  {journal} {Phys. Rev. B},\ }\textbf {\bibinfo {volume} {55}},\
  \bibinfo {pages} {9326} (\bibinfo {year} {1997})}\BibitemShut {NoStop}%
\bibitem [{Note5()}]{Note5}%
  \BibitemOpen
  \bibinfo {note} {According to Ref. \cite {Koulakov.PRL.1996,
  *Fogler.PRB.1997}, bubble phases with more than one electron per lattice site
  are not stable in the $N=0$ LL. (A bubble phase with one electron per lattice
  site is indistinguishable from a WC phase.)}\BibitemShut {NoStop}%
\bibitem [{\citenamefont {Suen}\ \emph {et~al.}(1994)\citenamefont {Suen},
  \citenamefont {Manoharan}, \citenamefont {Ying}, \citenamefont {Santos},\
  and\ \citenamefont {Shayegan}}]{Suen.PRL.1994}%
  \BibitemOpen
  \bibfield  {author} {\bibinfo {author} {\bibfnamefont {Y.~W.}\ \bibnamefont
  {Suen}}, \bibinfo {author} {\bibfnamefont {H.~C.}\ \bibnamefont {Manoharan}},
  \bibinfo {author} {\bibfnamefont {X.}~\bibnamefont {Ying}}, \bibinfo {author}
  {\bibfnamefont {M.~B.}\ \bibnamefont {Santos}}, \ and\ \bibinfo {author}
  {\bibfnamefont {M.}~\bibnamefont {Shayegan}},\ }\Doi
  {10.1103/PhysRevLett.72.3405} {\bibfield  {journal} {\bibinfo  {journal}
  {Phys. Rev. Lett.},\ }\textbf {\bibinfo {volume} {72}},\ \bibinfo {pages}
  {3405} (\bibinfo {year} {1994})}\BibitemShut {NoStop}%
\bibitem [{\citenamefont {Liu}\ \emph {et~al.}(2011)\citenamefont {Liu},
  \citenamefont {Shabani},\ and\ \citenamefont {Shayegan}}]{Liu.PRB.2011}%
  \BibitemOpen
  \bibfield  {author} {\bibinfo {author} {\bibfnamefont {Y.}~\bibnamefont
  {Liu}}, \bibinfo {author} {\bibfnamefont {J.}~\bibnamefont {Shabani}}, \ and\
  \bibinfo {author} {\bibfnamefont {M.}~\bibnamefont {Shayegan}},\ }\Doi
  {10.1103/PhysRevB.84.195303} {\bibfield  {journal} {\bibinfo  {journal}
  {Phys. Rev. B},\ }\textbf {\bibinfo {volume} {84}},\ \bibinfo {pages}
  {195303} (\bibinfo {year} {2011})}\BibitemShut {NoStop}%
\bibitem [{Note6()}]{Note6}%
  \BibitemOpen
  \bibinfo {note} {In all samples we measured, the $R_{xy}$ quantization is
  always more robust than the vanishing of $R_{xx}$. Similar behavior is seen
  for the RIQHE phases observed in the $N=1$ LL; see e.g. Ref. \cite
  {Eisenstein.PRL.2002}.}\BibitemShut {Stop}%
\bibitem [{Note7()}]{Note7}%
  \BibitemOpen
  \bibinfo {note} {W. Li \protect \emph {et al.} [Phys. Rev. Lett. {\protect
  \bf 105}, 076803 (2010)] have reported a RIQHE between filling factors $\nu
  =2/3$ and 3/5 and its particle-hole symmetric state, a reentrant insulating
  phase between $\nu =1/3$ and 2/5, in samples where 2D electrons reside in an
  Al$_x$Ga$_{1-x}$As alloy. They attribute their observation to the strong
  short-range (alloy) disorder which favors a pinned WC state in their
  samples.}\BibitemShut {Stop}%
\bibitem [{Note8()}]{Note8}%
  \BibitemOpen
  \bibinfo {note} {In the parameter range we study our samples, often two
  electric subbands, which are separated in energy by at least $\sim $40 K, are
  occupied. Near $\nu =1$, however, we expect that all the electrons occupy the
  lowest LL of the lowest electric subband. We therefore suspect the upper
  subband is not playing a major role, althought we cannot rule out the effect
  of subband mixing.}\BibitemShut {Stop}%
\bibitem [{\citenamefont {Price}\ \emph {et~al.}(1995)\citenamefont {Price},
  \citenamefont {Zhu}, \citenamefont {Das~Sarma},\ and\ \citenamefont
  {Platzman}}]{Price.PRB.1995}%
  \BibitemOpen
  \bibfield  {author} {\bibinfo {author} {\bibfnamefont {R.}~\bibnamefont
  {Price}}, \bibinfo {author} {\bibfnamefont {X.}~\bibnamefont {Zhu}}, \bibinfo
  {author} {\bibfnamefont {S.}~\bibnamefont {Das~Sarma}}, \ and\ \bibinfo
  {author} {\bibfnamefont {P.~M.}\ \bibnamefont {Platzman}},\ }\Doi
  {10.1103/PhysRevB.51.2017} {\bibfield  {journal} {\bibinfo  {journal} {Phys.
  Rev. B},\ }\textbf {\bibinfo {volume} {51}},\ \bibinfo {pages} {2017}
  (\bibinfo {year} {1995})}\BibitemShut {NoStop}%
\bibitem [{\citenamefont {Chen}\ \emph {et~al.}(2003)\citenamefont {Chen},
  \citenamefont {Lewis}, \citenamefont {Engel}, \citenamefont {Tsui},
  \citenamefont {Ye}, \citenamefont {Pfeiffer},\ and\ \citenamefont
  {West}}]{Chen.PRL.2003}%
  \BibitemOpen
  \bibfield  {author} {\bibinfo {author} {\bibfnamefont {Y.}~\bibnamefont
  {Chen}}, \bibinfo {author} {\bibfnamefont {R.~M.}\ \bibnamefont {Lewis}},
  \bibinfo {author} {\bibfnamefont {L.~W.}\ \bibnamefont {Engel}}, \bibinfo
  {author} {\bibfnamefont {D.~C.}\ \bibnamefont {Tsui}}, \bibinfo {author}
  {\bibfnamefont {P.~D.}\ \bibnamefont {Ye}}, \bibinfo {author} {\bibfnamefont
  {L.~N.}\ \bibnamefont {Pfeiffer}}, \ and\ \bibinfo {author} {\bibfnamefont
  {K.~W.}\ \bibnamefont {West}},\ }\Doi {10.1103/PhysRevLett.91.016801}
  {\bibfield  {journal} {\bibinfo  {journal} {Phys. Rev. Lett.},\ }\textbf
  {\bibinfo {volume} {91}},\ \bibinfo {pages} {016801} (\bibinfo {year}
  {2003})}\BibitemShut {NoStop}%
\bibitem [{\citenamefont {Chen}\ \emph {et~al.}(2004)\citenamefont {Chen},
  \citenamefont {Lewis}, \citenamefont {Engel}, \citenamefont {Tsui},
  \citenamefont {Ye}, \citenamefont {Wang}, \citenamefont {Pfeiffer},\ and\
  \citenamefont {West}}]{Chen.PRL.2004}%
  \BibitemOpen
  \bibfield  {author} {\bibinfo {author} {\bibfnamefont {Y.~P.}\ \bibnamefont
  {Chen}}, \bibinfo {author} {\bibfnamefont {R.~M.}\ \bibnamefont {Lewis}},
  \bibinfo {author} {\bibfnamefont {L.~W.}\ \bibnamefont {Engel}}, \bibinfo
  {author} {\bibfnamefont {D.~C.}\ \bibnamefont {Tsui}}, \bibinfo {author}
  {\bibfnamefont {P.~D.}\ \bibnamefont {Ye}}, \bibinfo {author} {\bibfnamefont
  {Z.~H.}\ \bibnamefont {Wang}}, \bibinfo {author} {\bibfnamefont {L.~N.}\
  \bibnamefont {Pfeiffer}}, \ and\ \bibinfo {author} {\bibfnamefont {K.~W.}\
  \bibnamefont {West}},\ }\Doi {10.1103/PhysRevLett.93.206805} {\bibfield
  {journal} {\bibinfo  {journal} {Phys. Rev. Lett.},\ }\textbf {\bibinfo
  {volume} {93}},\ \bibinfo {pages} {206805} (\bibinfo {year}
  {2004})}\BibitemShut {NoStop}%
\bibitem [{\citenamefont {Yi}\ and\ \citenamefont
  {Fertig}(1998)}]{Yi.PRB.1998}%
  \BibitemOpen
  \bibfield  {author} {\bibinfo {author} {\bibfnamefont {H.}~\bibnamefont
  {Yi}}\ and\ \bibinfo {author} {\bibfnamefont {H.~A.}\ \bibnamefont
  {Fertig}},\ }\Doi {10.1103/PhysRevB.58.4019} {\bibfield  {journal} {\bibinfo
  {journal} {Phys. Rev. B},\ }\textbf {\bibinfo {volume} {58}},\ \bibinfo
  {pages} {4019} (\bibinfo {year} {1998})}\BibitemShut {NoStop}%
\bibitem [{\citenamefont {Narevich}\ \emph {et~al.}(2001)\citenamefont
  {Narevich}, \citenamefont {Murthy},\ and\ \citenamefont
  {Fertig}}]{Narevich.PRB.2001}%
  \BibitemOpen
  \bibfield  {author} {\bibinfo {author} {\bibfnamefont {R.}~\bibnamefont
  {Narevich}}, \bibinfo {author} {\bibfnamefont {G.}~\bibnamefont {Murthy}}, \
  and\ \bibinfo {author} {\bibfnamefont {H.~A.}\ \bibnamefont {Fertig}},\ }\Doi
  {10.1103/PhysRevB.64.245326} {\bibfield  {journal} {\bibinfo  {journal}
  {Phys. Rev. B},\ }\textbf {\bibinfo {volume} {64}},\ \bibinfo {pages}
  {245326} (\bibinfo {year} {2001})}\BibitemShut {NoStop}%
\bibitem [{\citenamefont {Chang}\ \emph {et~al.}(2005)\citenamefont {Chang},
  \citenamefont {Jeon},\ and\ \citenamefont {Jain}}]{Chang.PRL.2005}%
  \BibitemOpen
  \bibfield  {author} {\bibinfo {author} {\bibfnamefont {C.-C.}\ \bibnamefont
  {Chang}}, \bibinfo {author} {\bibfnamefont {G.~S.}\ \bibnamefont {Jeon}}, \
  and\ \bibinfo {author} {\bibfnamefont {J.~K.}\ \bibnamefont {Jain}},\ }\Doi
  {10.1103/PhysRevLett.94.016809} {\bibfield  {journal} {\bibinfo  {journal}
  {Phys. Rev. Lett.},\ }\textbf {\bibinfo {volume} {94}},\ \bibinfo {pages}
  {016809} (\bibinfo {year} {2005})}\BibitemShut {NoStop}%
\bibitem [{\citenamefont {Chang}\ \emph {et~al.}(2006)\citenamefont {Chang},
  \citenamefont {T\"oke}, \citenamefont {Jeon},\ and\ \citenamefont
  {Jain}}]{Chang.PRB.2006}%
  \BibitemOpen
  \bibfield  {author} {\bibinfo {author} {\bibfnamefont {C.-C.}\ \bibnamefont
  {Chang}}, \bibinfo {author} {\bibfnamefont {C.}~\bibnamefont {T\"oke}},
  \bibinfo {author} {\bibfnamefont {G.~S.}\ \bibnamefont {Jeon}}, \ and\
  \bibinfo {author} {\bibfnamefont {J.~K.}\ \bibnamefont {Jain}},\ }\Doi
  {10.1103/PhysRevB.73.155323} {\bibfield  {journal} {\bibinfo  {journal}
  {Phys. Rev. B},\ }\textbf {\bibinfo {volume} {73}},\ \bibinfo {pages}
  {155323} (\bibinfo {year} {2006})}\BibitemShut {NoStop}%
\bibitem [{Note9()}]{Note9}%
  \BibitemOpen
  \bibinfo {note} {We observe a qualitatively similar development near $\nu
  =6/5$: The traces at $n>2.61\times 10^{11}$ cm${^{-2}}$ in Fig. 2 show the
  disappearance of the 6/5 FQHE as the RIQHE sets in, although we cannot reach
  high enough densities to see a clear reemergence of the 6/5 FQHE
  state.}\BibitemShut {Stop}%
\end{thebibliography}%

\end{document}